\documentclass{webofc}
\usepackage[varg]{txfonts}   
\usepackage{listings}
\usepackage{hyperref}
\begin{document}
\title{Multiwavelength Light Curve Parameters of Cepheid Variables}
%
%

\author{\firstname{Anupam} \lastname{Bhardwaj}\inst{1,2}\fnsep\thanks{\href{mailto:anupam.bhardwajj@gmail.com}{\tt anupam.bhardwajj@gmail.com}} \and
        \firstname{Shashi M.} \lastname{Kanbur}\inst{3} \and 
        \firstname{Marcella} \lastname{Marconi}\inst{4} \and 
        \firstname{Marina} \lastname{Rejkuba}\inst{2} \and 
        \firstname{Harinder P.} \lastname{Singh}\inst{1} \and 
        \firstname{Chow-Choong} \lastname{Ngeow}\inst{5}
}
\institute{Department of Physics \& Astrophysics, University of Delhi, Delhi 110007, India 
\and European Southern Observatory, Karl-Schwarzschild-Stra\ss e 2, 85748, Garching, Germany
\and State University of New York, Oswego, NY 13126, USA
\and INAF-Osservatorio astronomico di Capodimonte, Via Moiariello 16, 80131 Napoli, Italy
\and Graduate Institute of Astronomy, National Central University, Jhongli 32001, Taiwan
	  }

\abstract{%
We present a comparative analysis of theoretical and observed light curves of Cepheid variables using Fourier decomposition. The theoretical light curves at multiple wavelengths are generated using stellar pulsation models for chemical compositions representative of Cepheids in the Galaxy and Magellanic Clouds. The observed light curves at optical ({\it VI}), near-infrared ({\it JHK}$_s$) and mid-infrared (3.6 $\&$ 4.5-$\mu$m) bands are compiled from the literature. We discuss the variation of light curve parameters as a function of period, wavelength and metallicity. Theoretical and observed Fourier amplitude parameters decrease with increase in wavelength while the phase parameters increase with wavelength. We find that theoretical amplitude parameters obtained using canonical mass-luminosity levels exhibit a greater offset with respect to observations when compared to non-canonical relations. We also discuss the impact of variation in convective efficiency on the light curve structure of Cepheid variables. The increase in mixing length parameter results in a zero-point offset in bolometric mean magnitudes and reduces the systematic large difference in theoretical amplitudes with respect to observations. 
}
\maketitle

\section{Introduction}\label{sec:intro}

Classical Cepheids are radially pulsating variables that display periodic light curves and follow a well-defined Period-Luminosity relation \cite[PLR,][]{levitt12}. Cepheids play a vital role in extragalactic distance scale to determine an accurate and precise value of the Hubble Constant \cite{riess16}. In addition, a comparison of observed pulsation properties of these variables with theoretical pulsation models provides fundamental test of the theory of stellar pulsation and evolution. 

First quantitative study of light curves of Cepheid variables based on Fourier decomposition method was presented for a sample of 57 Cepheids \cite{slee81}. It was suggested that the lower order Fourier amplitude and phase coefficients contain the most characteristic features of the light curve structure \cite{simont82, kovacs86}. The Cepheid light curves display bump features as a function of period known as Hertzsprung progressions \cite[hereafter HP,][]{hertzsprung1926}. The variation of HP for Classical Cepheids was investigated using pulsation models and it was found that the central period of the progression occurs around 11 days for LMC Cepheids \cite{bono2000d}. The comparison of pulsation properties of Cepheids from theoretical models with observations in terms of PLR has been a subject of many studies in the past decade \cite{bono1999a, caputo2000b} but no rigorous comparison of light curve structure was carried out at multiple wavelengths for different compositions. 

The observed light curve structure of Cepheid variables in the Galaxy and LMC was studied in detail and the variation of light curve parameters as a function of period and wavelength was presented in \cite{bhardwaj2015}. This work was extended to theoretical models of Cepheid variables and a comparison was presented at multiple wavelengths \cite{bhardwaj2017}. We summarize the results from these analyses in the following sections. 

\section{Analysis and Results}\label{sec:results}

The theoretical light curves are obtained from full amplitude, nonlinear, convective hydrodynamical models and follow the same assumptions as discussed in \cite{marconi2013}. For a fixed chemical composition and mass, the luminosity levels are adopted from stellar evolutionary calculations, i.e., canonical mass-luminosity (M-L) relations. In addition, models with a brighter luminosity level by 0.25 dex (Non-canonical M-L relations) are also computed to account for possible overshooting and mass loss efficiency. The topology of the instability strip is explored for each combination of M-L levels and the bolometric light curves are obtained. These light curves are transformed to optical and near-infrared filters by empirical relations from static atmospheric models. The observed optical and infrared band light curves are compiled from several sources in literature and the details can be found in \cite{bhardwaj2015}.


The theoretical and observed light curves were fitted with a Fourier series in the following form - 

\begin{equation}
m = m_{0}+\sum_{k=1}^{N}A_{k} \sin(2 \pi k x + \phi_{k}),
\label{eq:foufit1}
\end{equation}

\noindent where, $x$ represents the phase along the pulsation cycle. The optimum order of fit ($N$) is obtained using the minimization of the successive residuals and the Fourier coefficients are used to formulate Fourier amplitude ratios and phase differences ($R_{k1} = A_k/A_1~\&~ \phi_{k1} = \phi_k - k\phi_1, ~\mathrm{for}~ k > 1$).

\begin{figure*}
\centering
  \begin{tabular}{@{}cc@{}}
    \includegraphics[width=6cm,clip]{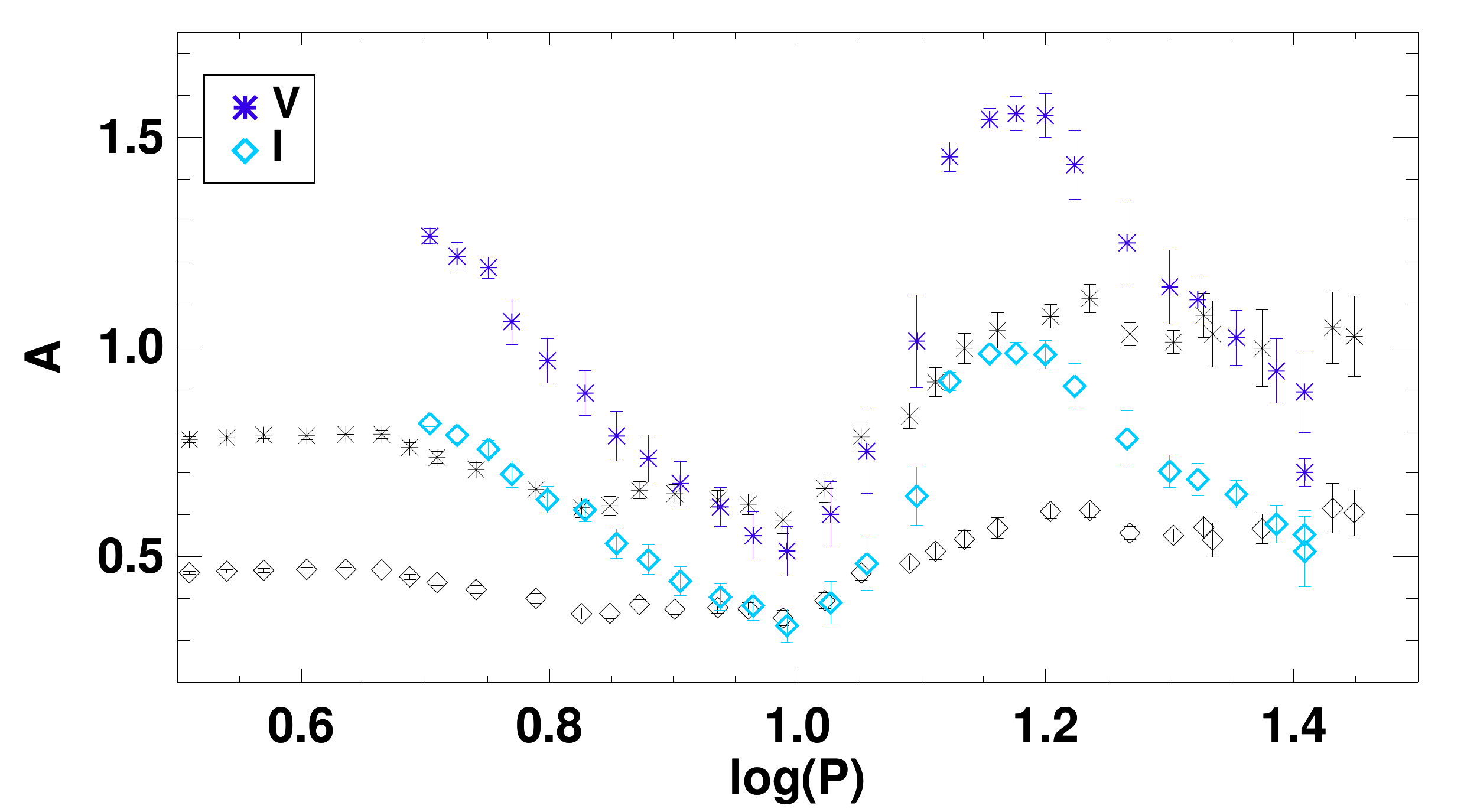} &  
    \includegraphics[width=6cm,clip]{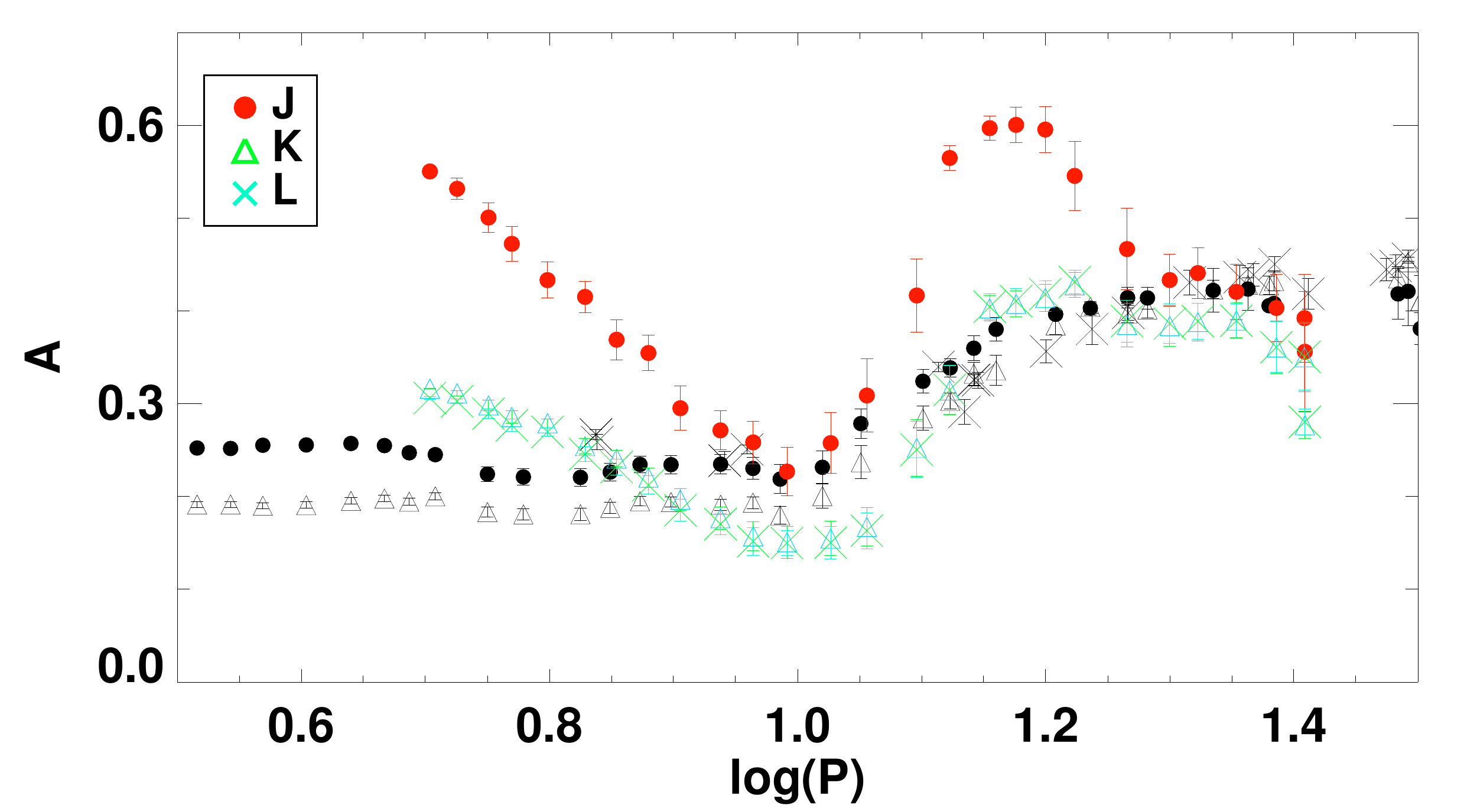}   \\
  \end{tabular}
 \caption{A comparison of observed (black symbols) and theoretical (colored symbols) mean amplitudes for fundamental-mode Cepheids in the LMC.}
 \label{fig:amp_multi}
\end{figure*}

The amplitude of the light curve is defined as the difference of maximum and minimum magnitudes obtained from the Fourier fit. Fig.~\ref{fig:amp_multi} displays the variation of mean amplitudes for the Cepheids in the LMC. The average values are obtained by sliding mean calculations in a bin-size of $\log(P)=0.1$~dex and in steps of 0.03~dex. The theoretical mean amplitudes are systematically larger than observed amplitudes at $VIJ$ wavelengths, as a function of period except close to the period of 10 days. At longer wavelengths ($KL$) the amplitudes are consistent between observations and theory. 

\subsection{Fourier parameters as a function of wavelength and metallicity}\label{subsec:fou_multi}

\begin{figure*}
\centering
  \begin{tabular}{@{}cc@{}}
    \includegraphics[width=6cm,clip]{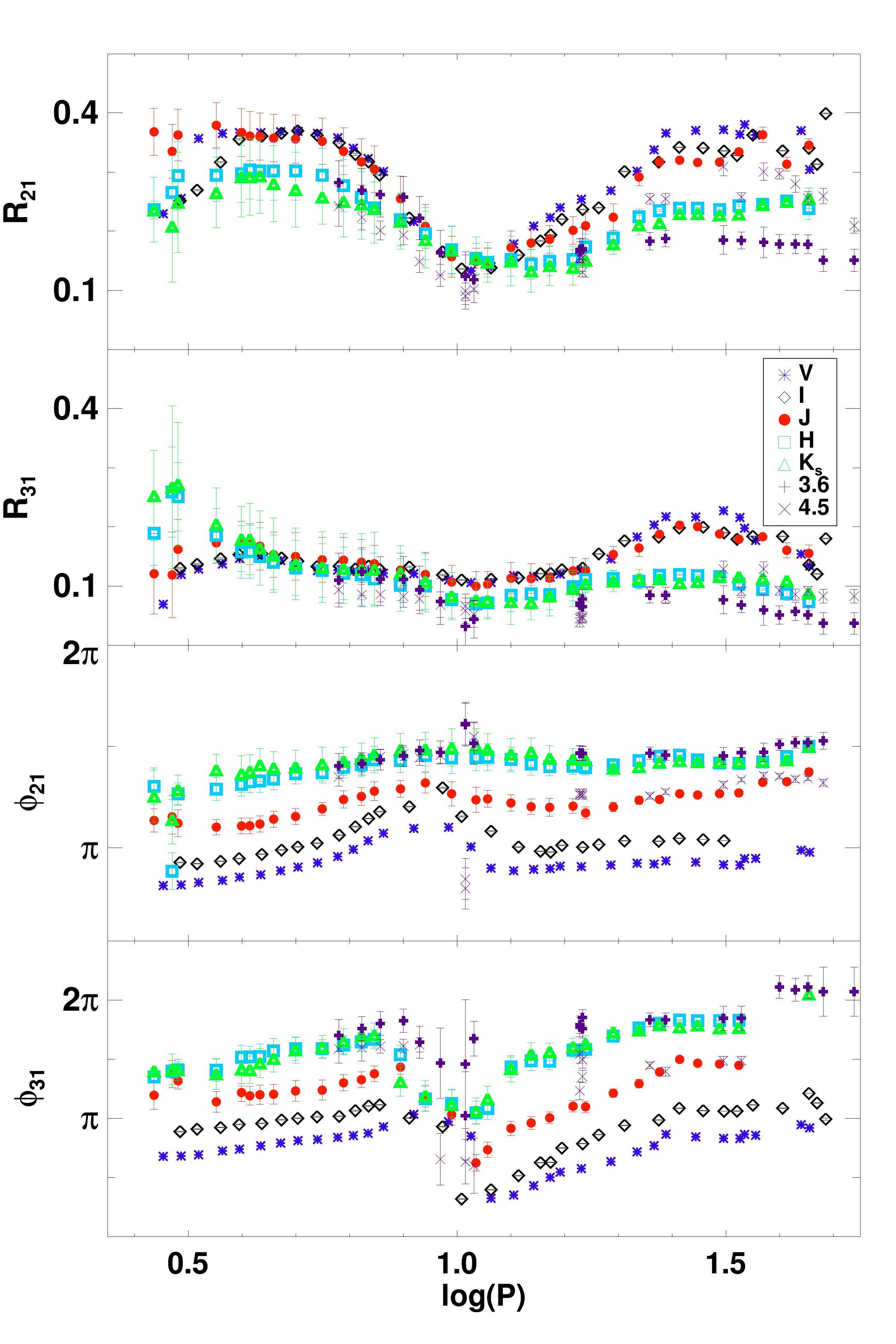} &  
    \includegraphics[width=6cm,clip]{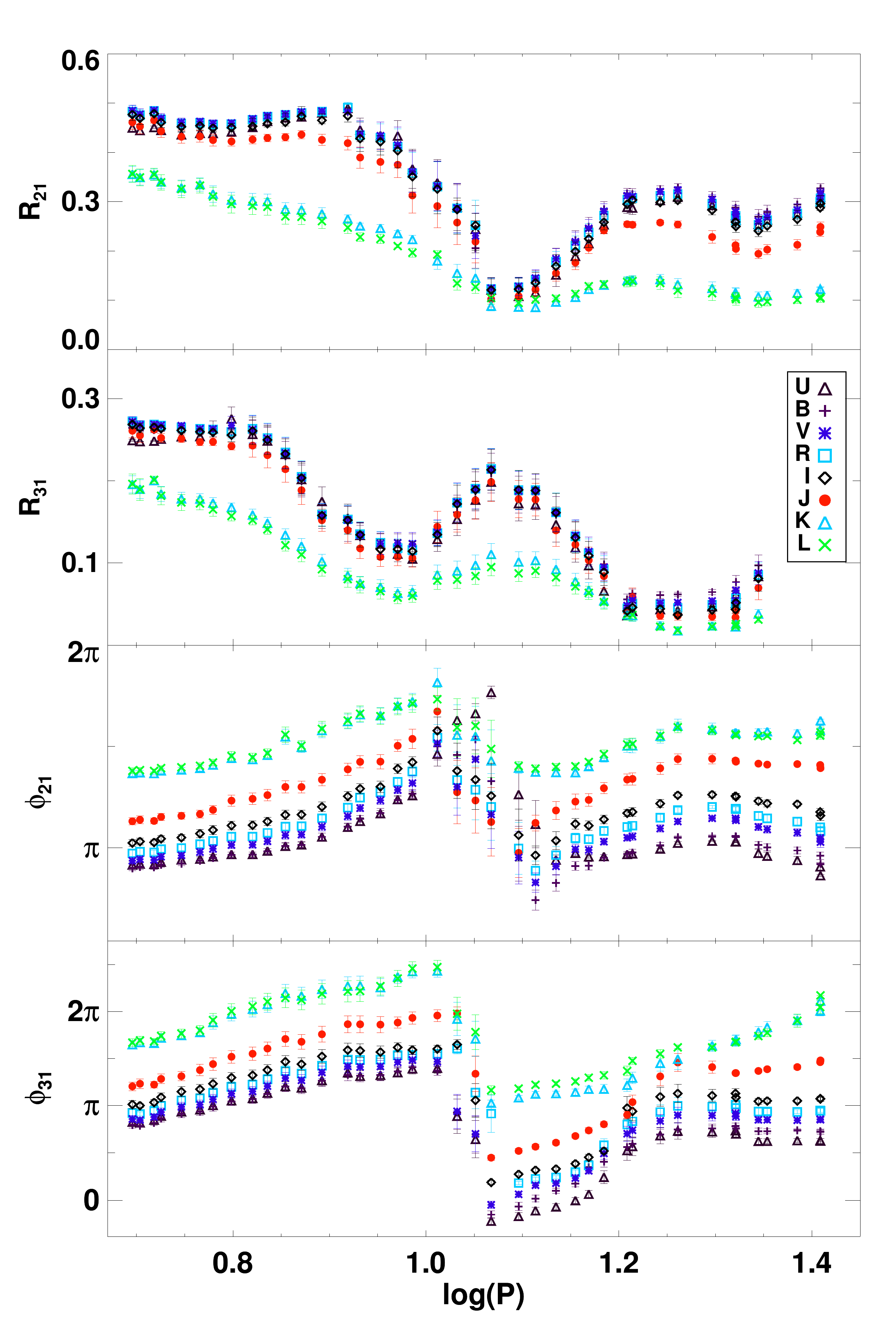}   \\
  \end{tabular}
 \caption{Multiband mean Fourier parameters for observed Galactic Cepheid light curves (left panel) and corresponding theoretical light curves (right panel) with chemical compositions, Y=0.28, Z=0.02.}
 \label{fig:fou_mean}
\end{figure*}

We discuss the variation of mean Fourier parameters as a function of period and wavelength for different metallicities. Fig.~\ref{fig:fou_mean} displays the variation of the mean Fourier parameters for Galactic Cepheids (left panels) and for theoretical models with metallicity, Z=0.02 (right panels) at multiple wavelengths. We find a decrease in Fourier amplitude parameters with increase in wavelength while the phase parameters increase with wavelength for both theoretical and observed light curves. In the long period range, $\log(P) > 1$, $R_{21}$ increases sharply to a peak value around $\log (P)=1.3$, at optical wavelengths. This variation is visible for both theoretical and observed Fourier parameters and separates optical wavelengths to redder bands. Similar features in Fourier parameters are also seen for Magellanic Cloud Cepheids \cite{bhardwaj2017}. The mean phase parameters display a systematic offset at all wavelengths for the Galaxy and Magellanic Cloud Cepheid models.

\begin{figure*}
\centering
\includegraphics[width=11cm,clip]{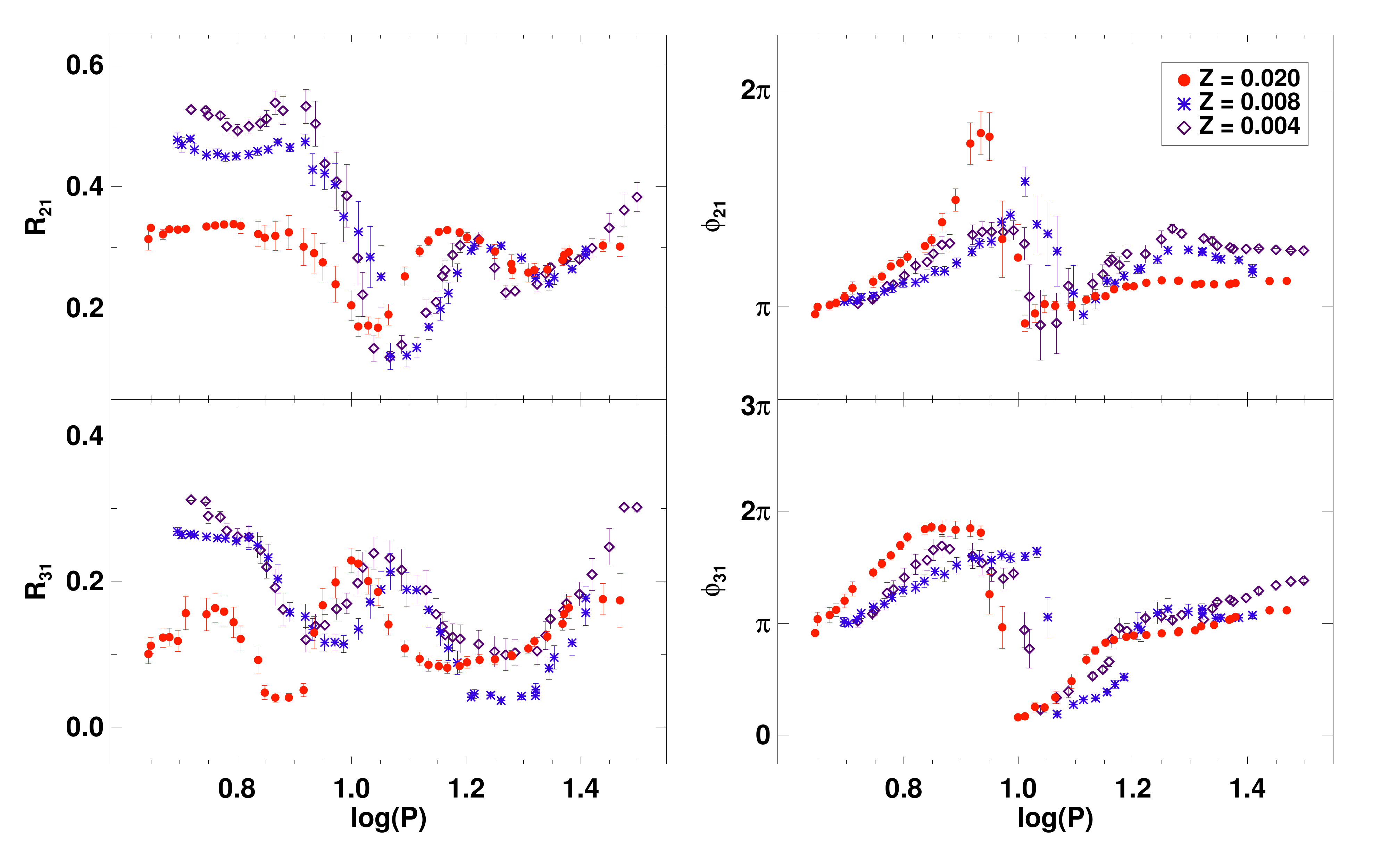}
\caption{Mean Fourier parameters in $I$-band for theoretical models of Cepheids with different metal abundances.}
\label{fig:fou_metal}       
\end{figure*}

Fig.~\ref{fig:fou_metal} displays variation in theoretical $I$-band mean Fourier parameters as a function of metallicity. The amplitude parameters in short period range ($\log(P) < 1$) increase with decrease in metallicity. The phase parameters also provide evidence of a decrease with metallicity but the progression is not distinctly clear. The central minimum in case of $R_{21}$ shifts to longer periods with decrease in metallicities. Similar variation in HP was also seen for observed light curves of Cepheids in \cite{bhardwaj2015}.

\subsection{A comparison of observed and theoretical Fourier parameters}\label{subsec:fou_comp}

\begin{figure*}
\centering
\includegraphics[width=11cm,clip]{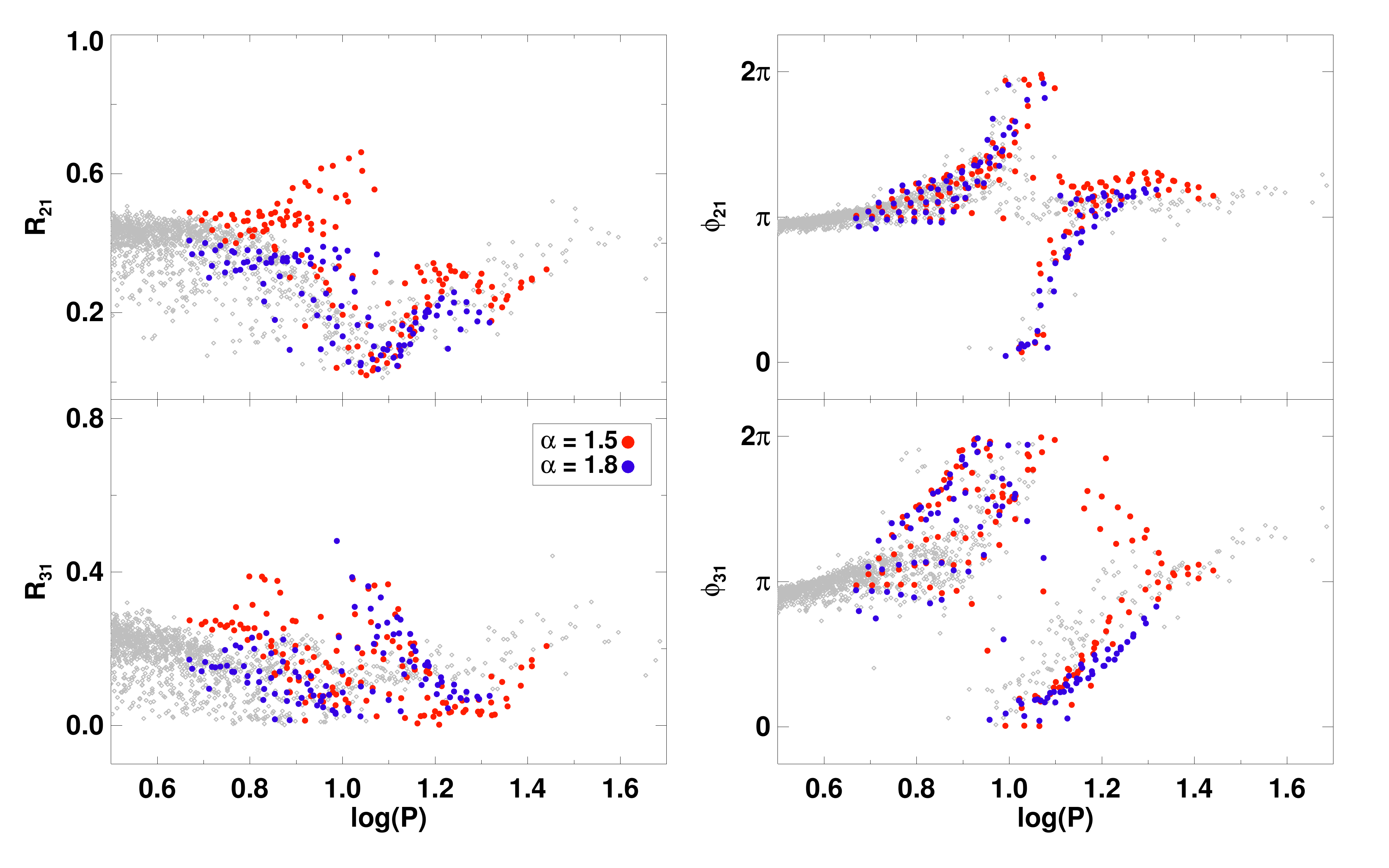}
\caption{Fourier parameters for theoretical and observed (grey) $I$-band light curves for Cepheids in the LMC. The theoretical models correspond to two mixing lengths, standard 1.5 (red) and enhanced 1.8 (blue).}
\label{fig:comp_fou}       
\end{figure*}

Fig.~\ref{fig:comp_fou} compares the Fourier parameters of the theoretical and observed $I$-band light curves for Cepheid variables in the LMC. The theoretical models are plotted for two different sets of mixing length parameters. For standard mixing length ($\alpha=1.5$), there is a large discrepancy in theoretical amplitude parameters at short period range ($\log(P) < 1$), when compared to observations. We find that canonical set of models displays larger offset in the overlapping period range. In contrast, the non-canonical Cepheid models are more consistent with observations. Increase in mixing length ($\alpha=1.8$) enhances the convective efficiency and damps down pulsation. In turn, it affects the width of the instability strip as the red edge becomes bluer. The increase in convection very likely leads to decrease in theoretical amplitudes. This is evident from Fig.~\ref{fig:comp_fou}, where increased mixing length (blue circles) helps in minimizing the discrepancy in theoretical amplitudes with respect to the observations. However, this also leads to zero-point offset in bolometric mean magnitudes. This may cause biases in distance estimates based on theoretical P-L relations with different mixing length parameters \cite{fiorentino2007}.

\section{Summary}

We discuss the variation of the light curve parameters as a function of period, wavelength and metallicity for the theoretical and observed light curves of Cepheid variables. We present a quantitative comparison between theoretical and observed light curve structure posing strong constraints for stellar pulsation codes that incorporate stellar atmosphere models to produce wavelength-dependent theoretical Cepheid light curves. We find a large discrepancy in theoretical amplitude parameters with respect to the observations and suggest that this difference can be reduced by adopting a combination of non-canonical set of models and increased mixing length. We note that the Fourier amplitude parameters can also be used to discriminate between canonical and non-canonical set of models in the overlapping period range. However, a more detailed analysis is required to further constrain the mass-luminosity relations obeyed by Cepheid variables and to explore the impact of different input parameters on the light curve structure. The mass-luminosity relations for Cepheids are based on stellar evolutionary calculations and therefore a thorough analysis with additional model computations on smoother grid, can potentially help constrain theories of stellar evolution and pulsation.

\begin{acknowledgement} 
\noindent\vskip 0.2cm
\noindent {\em Acknowledgments}: AB acknowledges the Senior Research Fellowship grant 09/045(1296)/2013-EMR-I from the Council of Scientific and Industrial Research (CSIR), India.
\end{acknowledgement}

\bibliographystyle{woc}
\bibliography{AB_ceph_SP}

%
%
%






\end{document}